\documentclass{aa}
\usepackage{graphicx}
\usepackage{txfonts}
\begin{document}

   \title{Dust distribution in circumstellar disks harboring multi-planet systems}

   \subtitle{I. Sub-thermal mass planets}

   \author{V. Roatti
          \inst{1}
          \and
          G. Picogna\inst{2}
          \and
          F. Marzari\inst{3}
          }

   \institute{Dipartimento di Fisica e Astronomia “G. Galilei”, Universit\`a degli Studi di Padova, Vicolo dell’Osservatorio 3, 35122 Padova, Italy\\
              \email{vincenzo.roatti@phd.unipd.it}
         \and
             Universit\"ats-Sternwarte, Ludwig-Maximilians-Universit\"at M\"unchen, Scheinerstr. 1, M\"unchen, 81679, Bayern, Germany\\
             \email{picogna@usm.lmu.de}
\and
Dipartimento di Fisica e Astronomia “G.Galilei”, Universit\`a degli Studi di Padova, Via Marzolo 8, 35121 Padova, Italy \\
\email{francesco.marzari@pd.infn.it}
             }

   \date{Received ... / Accepted ...}
 
  \abstract
   {}
   {We investigate the formation of dust gaps in circumstellar disks driven by the presence of multiple low-mass planets, focusing on the distinct physical mechanisms that operate across different gas-dust coupling regimes.}
   {We performed 2D hydrodynamical simulations of multiple planets embedded in a circumstellar disk using the PLUTO code, with the addition of dust treated as Lagrangian particles with a multi-size distribution. We carried out a large parameter space analysis to check the influence of disk and planetary properties on the dust component.}
   {Planets with $m \gtrsim 1  \, M_{\oplus}$ can open dust gaps for small grains in dense and warm disks (strong coupling) and for large grains in thin and cold disks (weak coupling), without significantly perturbing the gas. In the strong coupling regime, rapid Type I migration can shift the gap location inward or outward with respect to the planetary orbit, depending on the direction of migration. We also find dust gaps that overlap with Lindblad resonances. In the weak coupling regime, planets can create an inner dust cavity, multiple dust rings, or hide inside a common gap.}
   {Our results show how low-mass multi-planet systems perturb the dust distribution, which cannot be explained by considering each planet in isolation and has a crucial dependence on local disk conditions and dust grain sizes.}


   \maketitle
\section{Introduction}
High-resolution observations of millimeter dust continuum emission from circumstellar disks have revealed the ubiquitous presence of axisymmetric substructures in the dust distribution, such as gaps, rings, and cavities \citep{2015ApJ...808L...3A, 2018ApJ...869L..41A}. These features can be produced by a variety of physical processes that perturb the radial motion of dust grains. Those include zonal flows \citep{2015A&A...574A..68F, 2016A&A...589A..87B}, self-induced dust pileups \citep{2015MNRAS.454L..36G}, dust growth at snow lines \citep{2015ApJ...806L...7Z}, aggregate sintering \citep{2016ApJ...821...82O}, magnetic disk winds and accretion streams \citep{2017MNRAS.468.3850S}, radial pressure bumps \citep{2018ApJ...869L..46D}, large-scale instabilities due to dust settling \citep{2016MNRAS.457L..54L}, secular gravitational instabilities \citep{2014ApJ...794...55T, 2018PASJ...70....3T}, or thermal wave instabilities \citep{2008ApJ...672.1183W, 2021ApJ...914L..38U, 2021ApJ...923..123W}. \\ 
The other interpretation of dust substructures is related to embedded planets. These can open dust gaps by interacting with the surrounding gas \citep{2004A&A...425L...9P}. Specifically, analytical and numerical studies have shown that a giant planet embedded in a circumstellar disk excites spiral density waves at the location of Lindblad resonances, opening a gap in the surrounding gas across several Hill radii and increasing the local pressure gradient at the gap edges (see the review by \cite{2012ARA&A..50..211K} and references therein). This pressure bump acts as a trap for dust grains, stopping them from drifting toward the star, and therefore opening a gap in the dust distribution. This process is expected for super-thermal mass planets, where the thermal mass, M$_{\rm{th}}$, is defined as the planetary mass at which the Hill radius of the planet, $r_H$, is equal to the vertical scale height of the disk: $(M_{p} / M_{\star})_{th} = 3(H/r _{p})^3$, with $r_p$ the radius of the planet's orbit. \\
In recent work, it has been proposed that small planets of a few Earth masses (sub-thermal mass) could also open a dust gap, which is not necessarily correlated with gas pressure bumps. Two different physical mechanisms have been studied: first, \cite{2017MNRAS.469.1932D} considered the competition between the gap opening contribution of the tidal torque exerted by the planet and the gap-closing gas viscous torque. The balance between these two torques outside the planet's orbit determines whether the planet is able to carve a gap in the dust or not. This is important for dust grains that have a large Stokes number, meaning that they are less coupled to the gas and more prone to tidal interactions. In this case, the planet may open a gap in the dust without altering the gas distribution. Second, \cite{2022A&A...665A.122K} focus on dust grains with small Stokes numbers that are tightly coupled to the gas. In this context, the authors computed the radial gas outflow produced by a low-mass planet in the proximity of its orbit. Because dust grains are forced to follow the gas motion, the gas outflow deflects the grains' trajectories, inhibiting their radial inward drift. As a consequence, a dust gap is opened without any perturbation in the gas pressure gradient. \\
The efficiency of these processes critically depends on disk properties, including grain size distribution, disk temperature, gas density profiles, and the degree of coupling between dust and gas. In addition, several secondary effects can modulate the formation of gaps. These include: planetary migration, which can prevent gap formation \citep{2015ApJ...802...56M}; resonant interactions, which can enhance dust trapping \citep{2019AJ....157...45M}; planetary growth, which changes the torque balance over time \citep{2012ARA&A..50..211K}; and turbulent diffusion, which counteracts dust segregation \citep{2023MNRAS.520.2913M}. The presence of condensation fronts such as the snow line may further influence dust accumulation \citep{2015ApJ...806L...7Z}. Moreover, when multiple planets are present in the system, their combined gravitational influence on the disk and mutual gravitational interactions may alter the dust structure. As a consequence, the overall dust distribution may differ significantly from what would be expected by considering each planet in isolation. \\
In this paper, we investigate how low-mass planets embedded in typical circumstellar disks affect the surrounding dust, with a focus on identifying the dominant mechanisms responsible for dust gap opening across a range of particle sizes and Stokes numbers. For a given disk, all the gap opening mechanisms induced by the planets (gas outflows, tidal torques, and gas pressure bumps) may be at work simultaneously. Therefore, we aim to understand whether any observable features in the dust distribution can help to disentangle the three mechanisms.
We performed hydrodynamical simulations using the code PLUTO \citep{2007ApJS..170..228M}, modified to handle Lagrangian particles of varying sizes. We carried out a large parameter space analysis to check the influence of disk and planet parameters on the dust component.
\section{Numerical setup}
We simulated the evolution of two planets embedded in a gas disk using the hydrodynamical code PLUTO \citep{2007ApJS..170..228M}, which solves the Navier-Stokes equations on a static polar grid ($r, \phi$). We considered a thin, non-self-gravitating, locally isothermal viscous disk extending from 0.5 to 15 au, described by a grid with 342x682 elements, with a logarithmic spacing in radius and a uniform spacing in the azimuthal direction. This corresponds to a grid resolution of 4-6 cells per scale height (“cps”) at the planet's location, depending on the specific model considered. \\ 
For the initial conditions, we characterized the gas surface density in the disk using a power-law profile,
\begin{equation}
    \Sigma = \Sigma _0 \left( \frac{R}{1\, \mathrm{au}}\right)^{-p} \,,
\end{equation}
where $\Sigma _0$ and $p$ are initial parameters that are set at the beginning of the simulation. We chose different initial values in the range of $\Sigma _0 = $ 100 - 1000 g/cm$^2$ and $p = $ 0.1 - 1.0 to check the response of the dust to different disk masses. In fact, varying the gas density can change the regime of dust-gas interactions, which may impact the resulting dust substructures. The surface densities that we employed are lower than what is predicted for the minimum mass solar nebula \citep{1977MNRAS.180...57W, 1981PThPS..70...35H} to account for the time required by the planets to grow. During this time, viscous evolution and possibly photo-evaporation dissipate the gas disk \citep{2022EPJP..137.1132W}. \\
We assumed that the disk is locally isothermal, which implies that the disk temperature also follows a power-law function of the radius
\begin{equation}
    T = T_0 \left( \frac{R}{1\, \mathrm{au}}\right)^{-q} \,,
\end{equation}
with initial parameters $T_0$ and $q$. The locally isothermal approximation allowed us to ignore the energy equation and focus on a 2D framework. However, the same code could support the addition of heating and cooling processes in the context of 3D simulations (see, for example, \cite{2008A&A...488..429T}): we discuss potential limitations of our approach in Section \ref{caveats}. \\
We described the viscosity of the disk using the $\alpha$ prescription (\cite{1973A&A....24..337S}) with a turbulent viscosity parameter $\alpha$ = 0.004, which is the same value adopted by \cite{2017MNRAS.469.1932D}. We also performed simulations with $\alpha = 10^{-4}$, following the recent hypothesis that disk evolution may be driven by magnetic disk winds instead of viscosity (see \cite{2023NewAR..9601674R} and references therein). Although different values of the $\alpha$ parameter can significantly influence the evolution of the disk, the timescale of gap opening is much shorter than the evolution timescale, so we did not account for the evolution of the disk in the code. Instead of introducing the planets right at the start of the simulation, we gradually increased their mass from 0 to their final mass in a ramp-up time of 100 years, to avoid significant perturbations of the gas and the particles \citep{2006MNRAS.370..529D}. \\
We computed the acceleration due to the mutual gravitational attraction between the planets and the star as in \cite{2018A&A...616A..47T}. We smoothed the gravitational potential to avoid singularities in the numerical evaluation of the acceleration: in particular, we used a smoothing length of $\epsilon = 0.6$ H, as this value describes the vertically averaged forces very well \citep{2012A&A...541A.123M, 2025arXiv250904282C}. Moreover, we calculated the gravitational feedback of the disk onto the planets (and the star) to account for the planetary migration \citep{2018A&A...616A..47T}. \\
We modeled the solid component of the disk using a large number of Lagrangian particles representative of dust dynamics \citep{2018A&A...616A.116P}. We preferred this approach over the pressureless fluid because in this way, we could follow the trajectories of the individual particles. Moreover, the fluid approach is not particularly accurate when working with large particle sizes. We included 200000 particles with a multi-size distribution divided into 10 bins from 100 $\mu$m to 5.12 cm, separated by powers of 2. \\ 
Even if pebble-size particles are not typically observed in the millimeter range, pebble accretion is a crucial process in planet formation, and centimeter-size grains could represent the missing fraction of dust mass that is required for disks to have enough solid material to form planets (this is the known disk mass budget problem; see \cite{2018A&A...618L...3M} for reference). Moreover, the size of the dust grains has an important impact on the physical mechanism that acts in the gap opening process. On the one hand, larger particles are less coupled to the gas and more prone to tidal interactions, which could lead to a gap opening via planetary tidal torque \citep{2017MNRAS.469.1932D}. On the other hand, smaller grains are forced to follow the motion of the gas, and a gap could be opened by the radial gas outflow produced by the planet in its vicinity \citep{2022A&A...665A.122K}. By simulating a wide range of particle sizes at once, we could test both mechanisms at the same time and see which one is dominant. \\
At the start of the simulation, the dust grains had Keplerian velocities and were initially distributed to have the same number of particles at each radius, $r$, leading to a density distribution that decreases as $1/r$. This distribution could be re-normalized to obtain any radial density profile for the dust. Note that we did not include dust growth and fragmentation in our code, as it would be numerically challenging: we discuss the possible consequences of omitting these processes in Section \ref{caveats}. \\ 
We integrated the trajectories of the dust particles together with the hydrodynamical evolution of the gas using a semi-implicit scheme, following \cite{2014ApJ...785..122Z}. In particular, we computed the forces acting on a dust particle: the gravitational acceleration due to the planets and the star, and the gas drag force. \\ 
Because of pressure support, the gas rotates at a slightly sub-Keplerian velocity. As a consequence, the dust particles experience a headwind, causing them to drift radially toward the star. Drag forces can be calculated in different regimes depending on the relationship between the mean free path of the gas, $\lambda$, and the particle size, $s$, which can be expressed in terms of the Knudsen number, Kn = $\lambda / 2s$. High Knudsen numbers correspond to strong dust-gas coupling in the Epstein regime, whereas Stokes drag applies at low Knudsen numbers. Because we considered a wide range of grain sizes, we computed the drag coefficient by interpolating the Epstein and Stokes regimes as in \cite{2009A&A...493.1125L},
\begin{equation} \label{Cd}
    C_D = \frac{9\mathrm{Kn}^2 C _D ^{\mathrm{Eps}} +C _D ^{\mathrm{Stk}} }{(3\mathrm{Kn} + 1)^2}\,,
\end{equation}
where $C _D ^{\mathrm{Eps}}$ and $C _D ^{\mathrm{Stk}}$ are the coefficients of Epstein and Stokes drag, respectively. From equation \ref{Cd} we can compute the dust stopping time,
\begin{equation}
    \tau _s = \frac{4 \lambda \rho _d}{3 \rho _g C_D c_s}\frac{1}{\mathrm{MaKn}}\,,
\end{equation}
where $\rho _d$ = 2 g/cm$^3$ is the material density of dust particles, $\rho _g$ is the gas density, $c_s$ is the sound speed and $Ma = v/c_s $ is the Mach number. Then, the drag force acting on a dust grain moving with velocity $\mathbf{v}_{\mathrm{rel}}$ relative to the gas is simply given by
\begin{equation}
    F_D = - \frac{\mathbf{v}_{\mathrm{rel}}}{\tau _s} .
\end{equation}
Because of gas drag, particles tend to disappear from the simulation when they reach the central star. Therefore, to study the long term evolution of the system, we fixed the number of dust particles in the outermost ring at every timestep, and thus kept a constant dust flux through the disk. \\
In addition to drag forces, we computed the gravitational acceleration onto the particles due to the star and the planet(s), while the back-reaction of the dust is ignored. Moreover, we accounted for the diffusion of dust grains due to turbulent motions by adding kicks in the positions of the particles, as in \cite{2011ApJ...737...33C}. Finally, when the grains reach the water snowline, we halved their size to account for the evaporation of the ices, and reduced their mean densities. We computed the location of the water snowline as the radius at which the pressure of the water vapor is equal to the equilibrium pressure (see \cite{2020A&A...635A.149G}, eq. 34).
\section{Results}
\begin{table}
\caption{Model parameters used in the simulations.}
\label{tab:model_parameters}
\centering
\resizebox{\linewidth}{!}{
\begin{tabular}{ c  c  c  c  c  c  c  c }
\hline\hline
{Models} & {1} & {2} & {3} & {4} & {5} & {2S} & {5S}  \\
\hline
$m_1$ (M$_{\oplus}$)           & 5   & 1  & 10  & 10 & 10 & 1 & 10 \\
$m_2$ (M$_{\oplus}$)           & 1   & 1  & 5   & 5 & 5 & \slash & \slash \\
$a_1$ (au)                     & 2   & 1  & 4   & 4 & 5 & 1 & 5\\
$a_2$ (au)                     & 3.3 & 1.58 & 6.5 & 6.5 & 10 & \slash & \slash \\
$\Sigma_0$ (g/cm$^2$)          & 1000 & 1000 & 100 & 100 & 0.1 & 1000 & 0.1 \\
$p$                            & 0.5 & 1.0 & 1.0 & 1.0 & 0.1 & 1.0 & 0.1\\
$T_0$ (K)                      & 300 & 200 & 200 & 200 & 300 & 200 & 300\\
$q$                            & 0.5 & 0.9 & 1.0 & 1.0 & 0.5 & 0.9 & 0.5\\
$\alpha$                       & 0.004 & 0.004 & 0.004 & $10^{-4}$ & 0.004 & 0.004 & 0.004 \\
\hline
\end{tabular}}
\end{table}
\begin{figure}
    \centering
    \includegraphics[width=\linewidth]{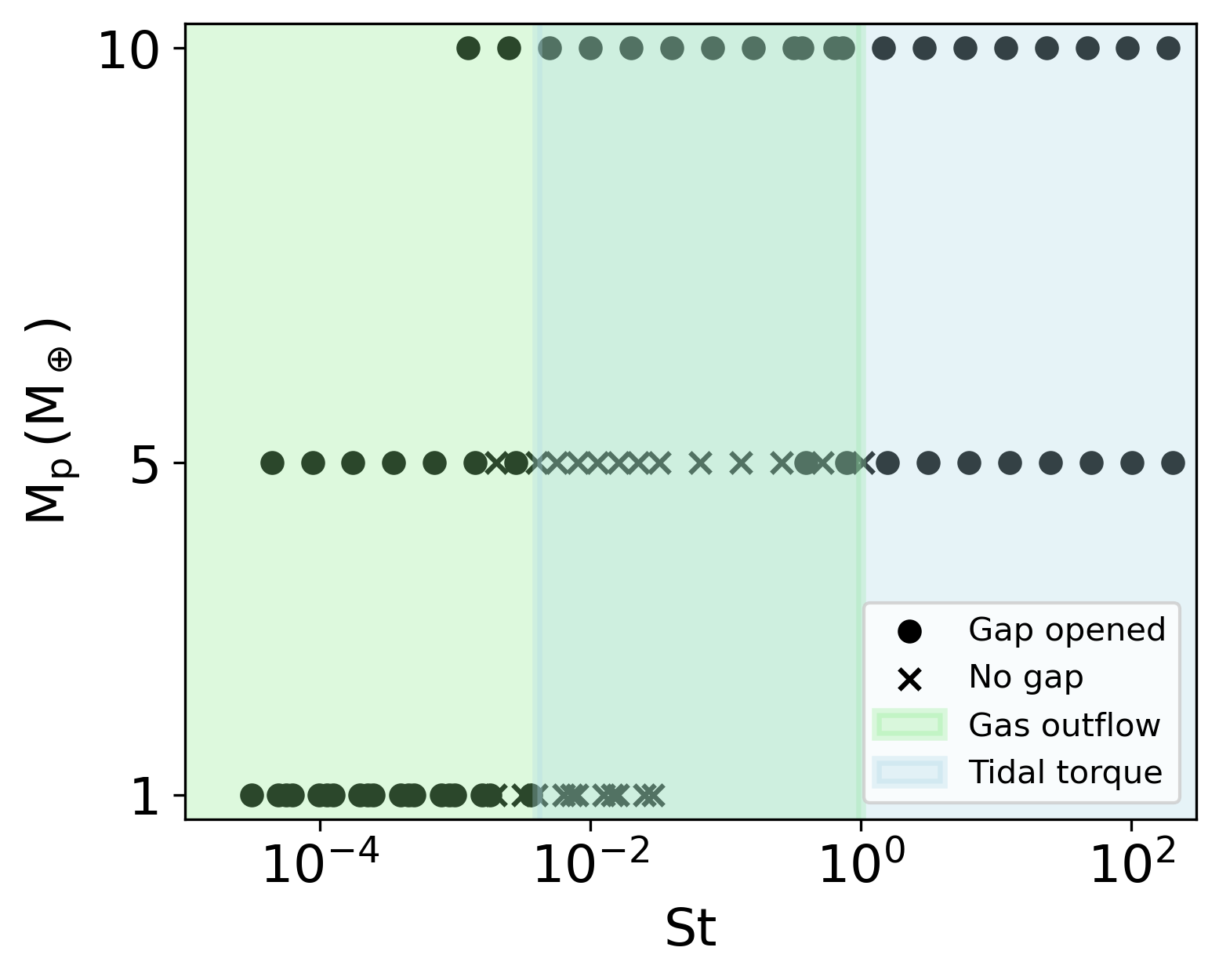}
    \caption{Schematic representation of the proposed dominant gap opening mechanism for a sub-thermal mass planet, as a function of planetary mass and Stokes number of the dust particles. Black dots (crosses) indicate the dust sizes considered in our models for which we found (no) dust gaps.}
    \label{scheme}
\end{figure}
In the following section, we show the results of different models that we considered by varying the physical properties of the disk and/or the masses and semimajor axes of the embedded planets. Table \ref{tab:model_parameters} resumes all the model parameters used in the simulations. Figure \ref{scheme} schematically represents the proposed dominant gap opening mechanism for a sub-thermal mass planet (M$_p$ < M$_{\mathrm{th}}$), as a function of the planetary mass and Stokes number of the dust particles. The superpositions of the colored regions highlight the fact that the boundaries of each mechanism are not rigorously defined.
\subsection{Models with low Stokes numbers}
\begin{figure*}[h!]
    \centering
    \includegraphics[width = \linewidth]{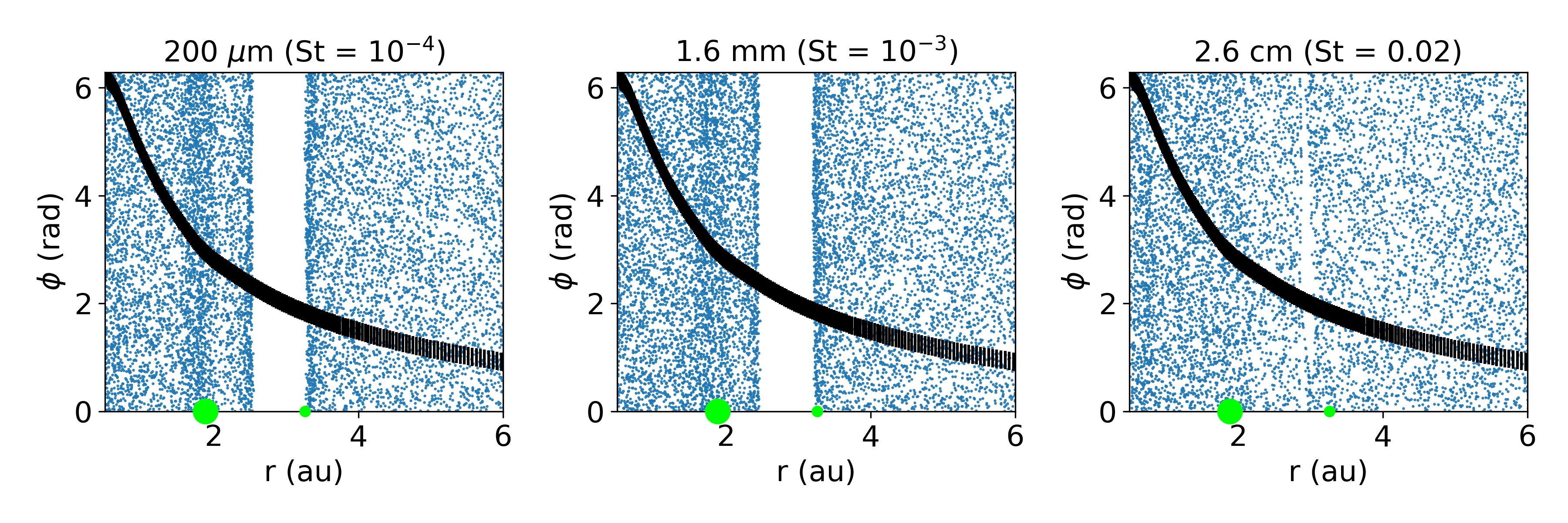}
    \caption{Dust distribution in Model 1. From left to right, the dust particle distributions in the $(r, \phi)$ plane for 200 $\mu$m, 1.6 mm, and 2.6 cm dust sizes in Model 1, respectively. The black lines show radial density distribution of the gas normalized between (0, 2$\pi$), while the filled green circles mark the position of the planets. The plots are zoomed in the inner region (0.5,6) au and the width of the density profile is multiplied by a constant for readability.}
    \label{model1_dust}
\end{figure*}
\begin{figure}[h!]
    \centering
    \includegraphics[width=\linewidth]{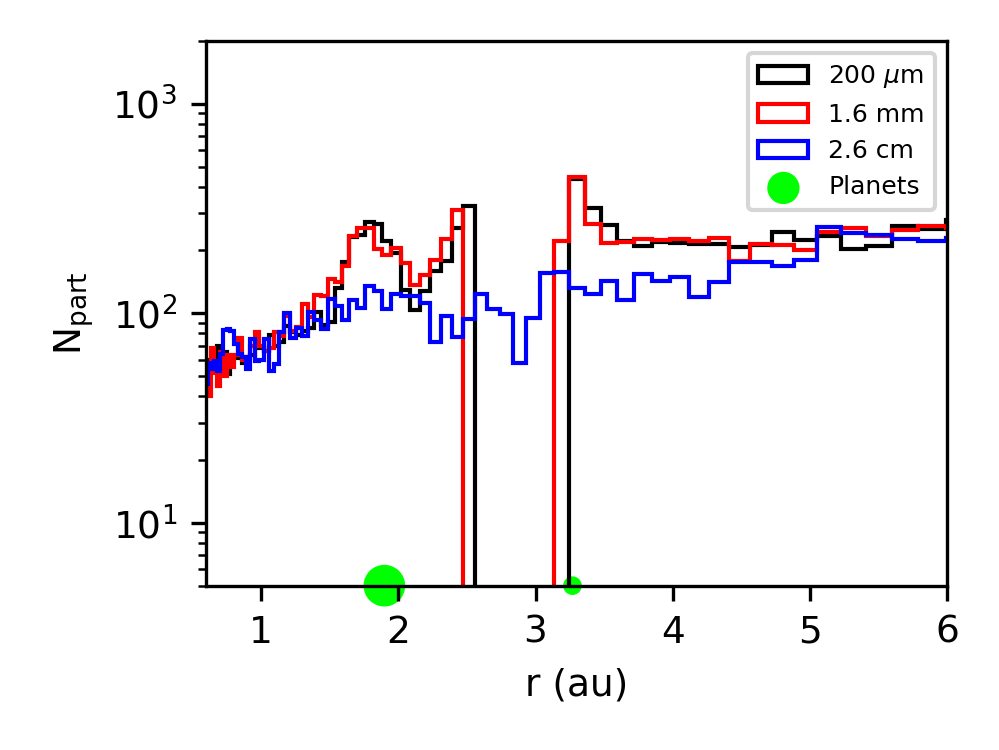}
    \caption{Histogram illustrating the number density of dust particles in Model 1 as a function of radial distance from the star. The number density is computed in 100 radial bins with logarithmic spacing. Filled green circles mark the positions of the planets. The black line represents 200-micron particles, while the red and blue lines represent 1.6-mm and 2.6-cm grains, respectively.}
    \label{model1_hist}
\end{figure}
\begin{figure*}[h!]
    \centering
    \includegraphics[width = \linewidth]{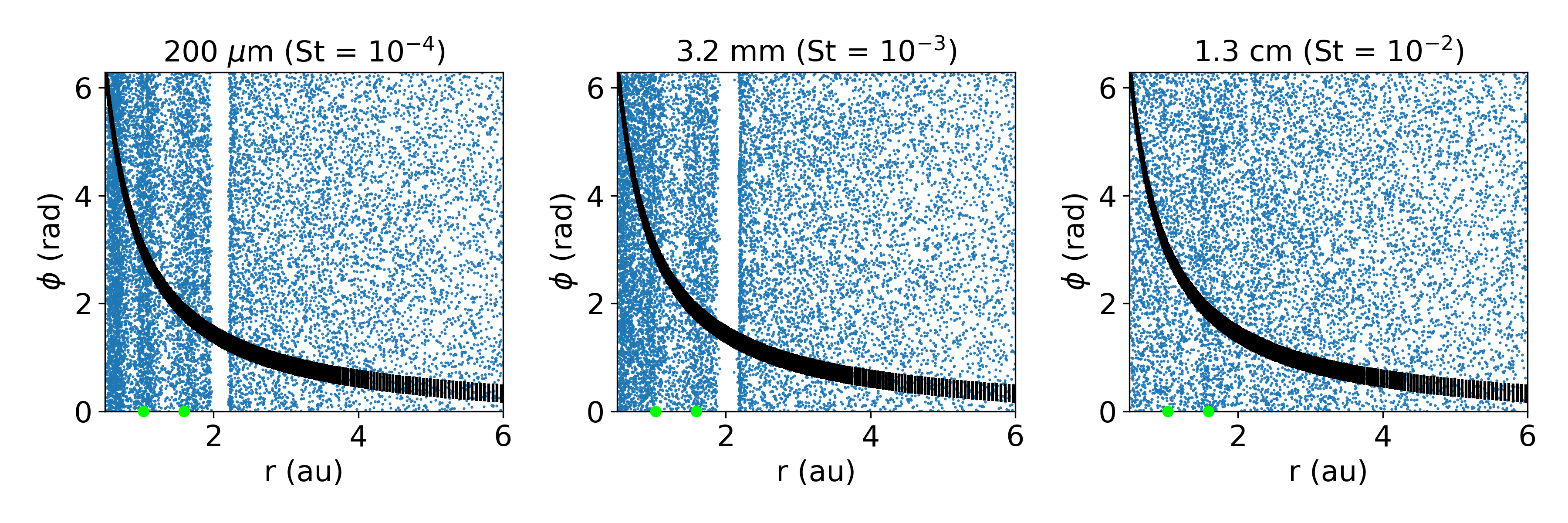}
    \caption{Dust distribution in Model 2. From left to right, the dust particle distributions in the $(r, \phi)$ plane for 200 $\mu$m, 3.2 mm, and 1.3 cm dust sizes in Model 2, respectively. The black lines show radial density distribution of the gas normalized between (0, 2$\pi$), while the filled green circles mark the position of the planets. The plots are zoomed in the inner region (0.5,5) au and the width of the density profile is multiplied by a constant for readability.}
    \label{model2_dust}
\end{figure*}
\begin{figure}[h!]
    \centering
    \includegraphics[width=\linewidth]{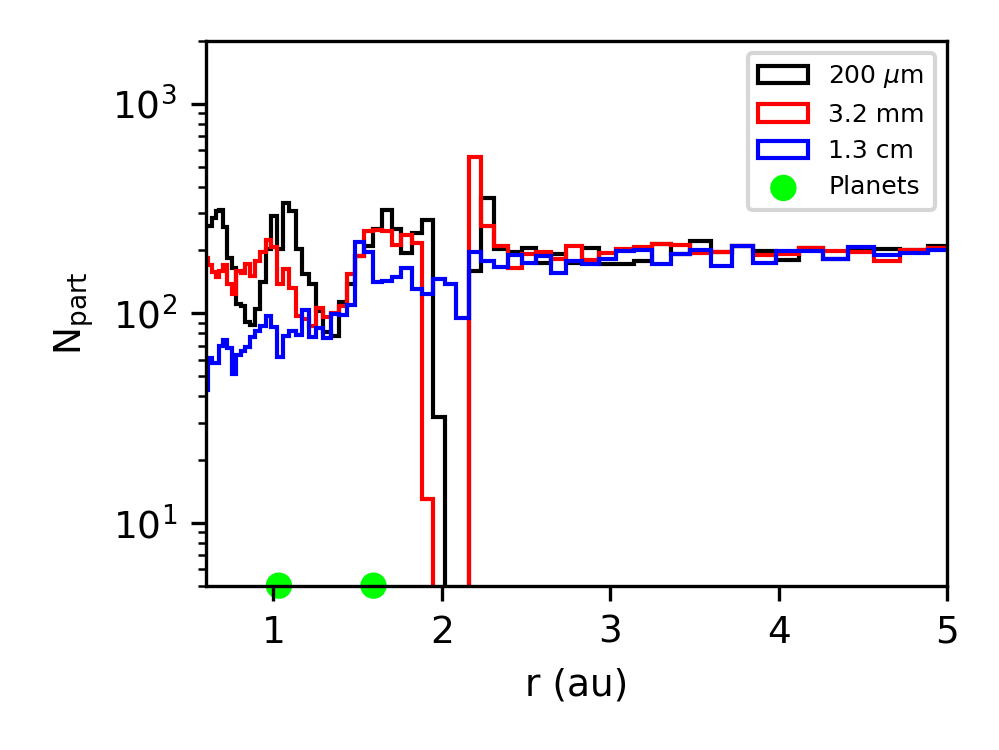}
    \caption{Same as Fig. \ref{model1_hist}, but for Model 2. The black line represents 200-micron particles, while the red and blue lines represent 3.2-mm and 1.3-cm grains, respectively.}
    \label{model2_hist}
\end{figure}
In Model 1, we examined a massive, somewhat evolved protoplanetary disk characterized by a high initial gas surface density of $\Sigma_0 = 1000$ g/cm$^2$ at 1 au, with index $p = 0.5$. The disk temperature profile was modeled with $T_0 = 300$ K and index $q = 0.5$. \\
To explore the interaction between planets and dust in this regime, we focused on dust particles with low Stokes numbers (St = $\tau_s \Omega _K$, with $\tau _s$ the stopping time of the dust particles and $\Omega _K$ the Keplerian frequency), which means that they are tightly coupled to the gas. This condition is necessary to test the gas outflow-driven gap opening mechanism proposed by \citet{2022A&A...665A.122K}, which predicts dust depletion without requiring significant perturbations of the gas. \\
We introduced two low-mass planets in the disk, with masses of $m_1 = 5 \, M_\oplus$ and $m_2 = 1 \, M_\oplus$, initially located near the 2:1 mean motion resonance at semimajor axes of $a_1 = 2$ au and $a_2 = 3.3$ au, respectively. However, due to the high gas surface density, both planets experienced rapid Type I migration, leading to divergent migration that quickly separated them from the resonant configuration (see Fig. \ref{orbital_evolution}). \\
Figure \ref{model1_dust} displays the resulting spatial distributions of gas and dust for three representative grain sizes. Interestingly, we find that a dust gap is formed only for grains of sizes $s \leq 6.4$ mm (corresponding to Stokes numbers $\mathrm{St} < 10^{-2}$), while the gas profile remains essentially unperturbed (as is seen from the black profiles in Fig. \ref{model1_dust}). This is consistent with the outflow-driven mechanism described by \citet{2022A&A...665A.122K}, where small grains are removed by gas outflows launched near the planet, even in the absence of a significant gas gap. Our results suggest that even an Earth-mass planet can carve a localized dust gap under these conditions, provided that the grains are well coupled to the gas. \\
Although the gap in the total dust surface density is shallow, we observe a nearly complete depletion of small grains in a narrow region just inside the orbit of the outer planet. In particular, this location differs from the prediction of \citet{2022A&A...665A.122K}, where the gas outflow induced by the planet causes dust depletion around the planetary orbit. The observed offset may result from the combined effects of planet migration and mutual gravitational interactions between the two planets. \\ 
To quantify the substructure, we show in Figure \ref{model1_hist} the number of particles as a function of the radial distance for three different dust sizes (see also the first panel of Fig. \ref{gapwidths}). The observed gap in the small grains is narrow ($\Delta _{\mathrm{gap}} \simeq 0.8$ au), but deep, since all the grains coming from the outer disk have stopped at the outer edge of the gap.
This result highlights the complexity of interpreting observations of protoplanetary disks. A dust gap does not necessarily reveal the presence of a single planet inside it, as multiple bodies and their dynamical evolution can alter the shape and location of dust substructures. Therefore, caution is warranted when attempting to infer planet masses from dust continuum observations alone. \\
\begin{figure}[h]
    \centering
    \includegraphics[width=\linewidth]{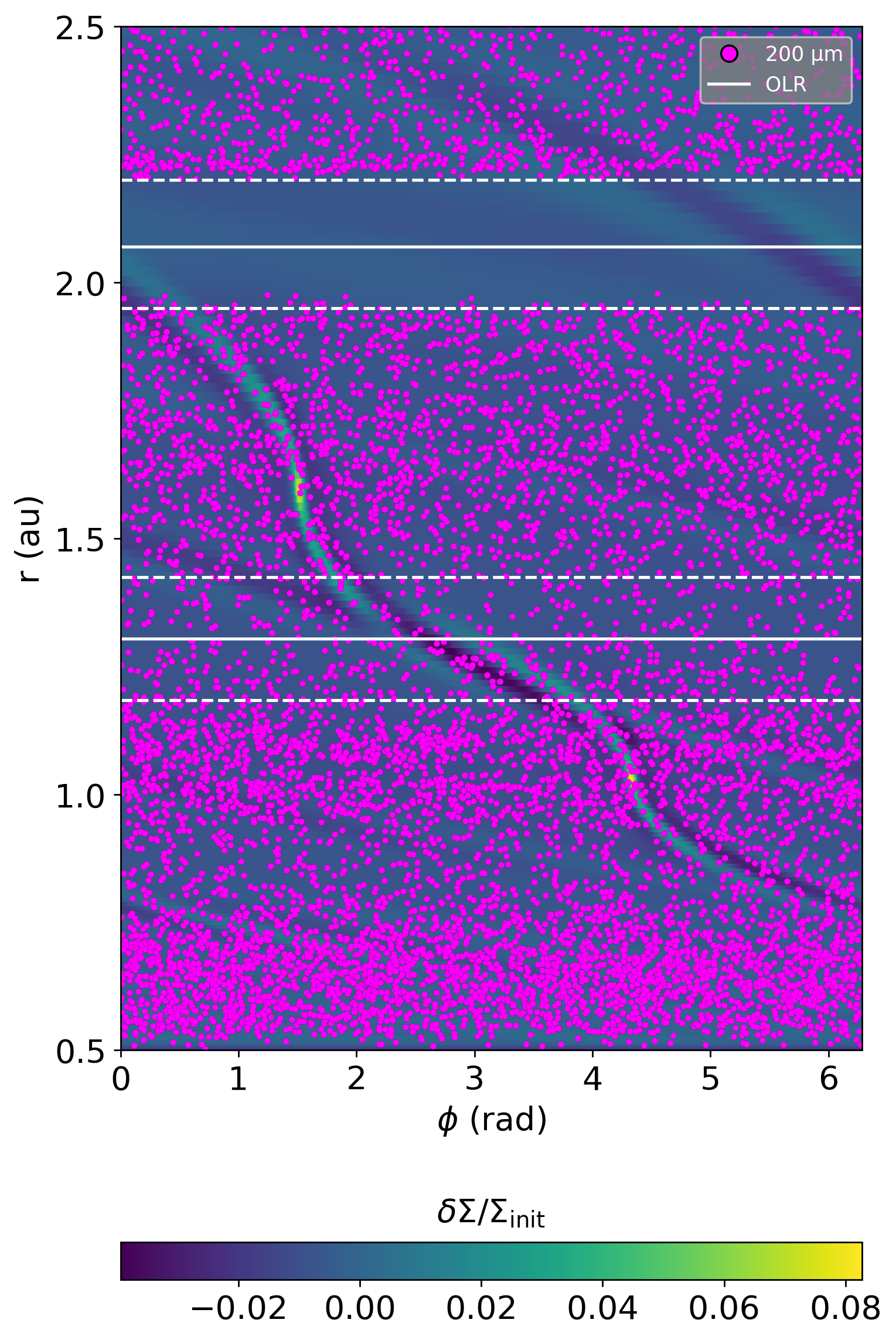}
    \caption{2D distributions of the perturbed gas surface density $\delta \Sigma /\Sigma _{\rm{init}}$ in Model 2, overlapped with the dust distribution of 200 $\mu$m grains. Solid white lines mark the location of the second-order OLR for the two planets. Dashed white lines mark the borders of the dust gap and the dust depletion discussed in the main text.}
    \label{model2_sig}
\end{figure}
In Model 2, we modified the thermal structure of the disk to more closely match the conditions explored by \citet{2022A&A...665A.122K}. Specifically, we reduced the disk temperature to $T_0 = 200$ K at 1 au and steepened the radial temperature gradient with a power law index $q = 0.9$. These changes affect both the gas scale height and the coupling between dust and gas, as the Stokes number increases with decreasing temperature for a given grain size. \\
We considered two equal-mass planets with masses of $m_1 = m_2 = 1 \, M_\oplus$, located near the 2:1 mean motion resonance at the initial semimajor axes of $a_1 = 1$ au and $a_2 = 1.58$ au. Despite the convergent nature of their migration in this case, both planets again failed to become trapped in resonance (see Fig. \ref{orbital_evolution}). This suggests that at the assumed gas density and viscosity, the migration timescale remains too short for efficient resonance capture, especially for Earth-mass planets. \\
Figure \ref{model2_dust} presents the gas and dust distribution resulting from this model for three different grain sizes. A clear dust gap forms for particles with sizes $s \leq 3.2$ mm, which is smaller than in the previous case due to the higher Stokes numbers and lower planetary masses. Still, as in the previous model, the gas distribution remains smooth, reinforcing the conclusion that the dust gap originates from the dust-gas interaction influenced by planet-induced gas flows. Interestingly, the dust gap in Model 2 is centered near 2 au, significantly beyond the orbit of the outer planet. This contrasts with Model 1, where the gap formed slightly interior to the orbit of the outer planet. The observed shift in gap location may be attributed to the direction of planetary migration: in Model 2, both planets migrate outward, because of the steeper temperature gradient and the symmetric mass configuration. The direction of migration may affect where the gas outflows accumulate, and consequently the dust depletion, thereby displacing the location of the gap from the planetary orbits. \\
In addition to the main dust gap at 2 au, we observe minor depletions in dust density between the two planets. This can be seen in Figure \ref{model2_hist}, where the decrease in the density of small grains around 0.8 and 1.3 au is less than an order of magnitude, while the gap observed around 2 au is narrow ($\Delta _{\mathrm{gap}} \simeq $ 0.3 au), but deep. The minor depletions in dust density may result from mutual gravitational perturbations between the planets, which could be more significant here because of their smaller orbital separation compared to the other models.
The location of the dust gap at 2 au and the dust depletion at 1.3 au may also be related to Lindblad resonances. These are orbital resonances around which a planet excites spiral density waves that propagate in the disk. Specifically, density waves are launched around the Lindblad resonance at \citep{1979ApJ...233..857G}
\begin{equation}
    r_m = \left( 1 \pm \frac{1}{m} \right) ^{\frac{2}{3}}r_{p},
\end{equation}
where $r_p$ is the radius of the circular orbit of the planet and $m$ is the azimuthal wavenumber of the resonance. These density waves propagate outward and inward in the disk. Interestingly, the dust gap observed at 2 au coincides with the location of the second-order ($m = 2$) outer Lindblad resonance (OLR) for the outer planet, while the dust depletion observed at 1.3 au corresponds to the second-order Lindblad resonance for the inner planet. This can be seen in Figure \ref{model2_sig}, where we plot the dust distribution of 200 $\mu$m size grains, overlapped with the perturbed surface density $\delta \Sigma /\Sigma _{\rm{init}}$, with $\delta \Sigma = \Sigma - \Sigma _{\rm{init}}$ and $\Sigma _{\rm{init}}$ the initial unperturbed gas surface density. The reason why we observe a gap in the dust distribution around 2 au might be due to the OLR from the outer planet with $m=2$ (2.07 au, solid white line in Figure \ref{model2_sig}) which strongly affects the gas and thus the dynamically coupled dust, opening a gap over time. The depletion of dust at 1.3 au could then be caused by the competition between the OLR with $m = 2$ (1.31 au, solid white line in Figure \ref{model2_sig}) from the inner planet and the inner Lindblad resonance with $m = 4$ (1.304 au) from the outer one. Although less significant, the small dust depletion within 0.8 au could also be related to the inner Lindblad resonance with $m=4$ (0.85 au) from the inner planet. This overlap between dust features and Lindblad resonances is observed only in this configuration, but it would imply that planetary torques can affect the small dust grains because of their strong coupling to the gas. \\
Together, the results of Models 1 and 2 highlight the sensitivity of dust gap formation to both disk thermodynamics and the architecture of the planetary system. In particular, temperature gradients and planet mass ratios play a critical role in determining the location and depth of dust gaps. This underscores the challenge of interpreting observed disk structures, which may reflect a complex interplay of disk physics and multi-body dynamics.
\subsection{Models with high Stokes numbers}
\begin{figure*}[h]
    \centering
    \includegraphics[width=\linewidth]{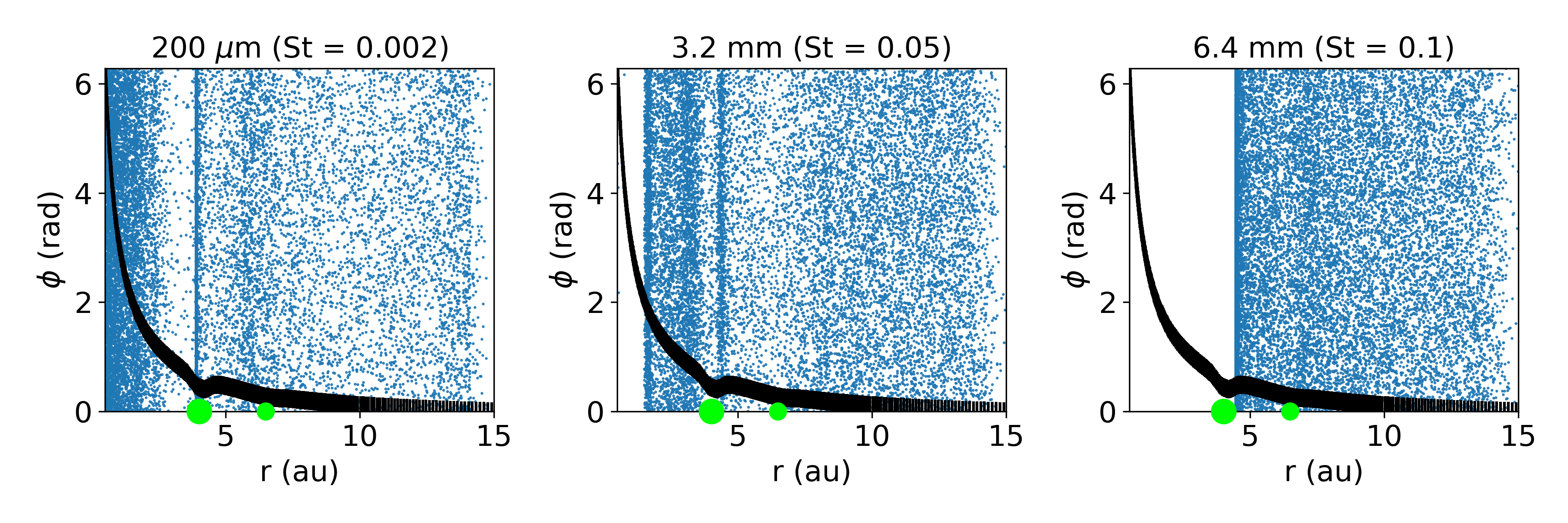}
    \caption{Dust distribution in Model 3. From left to right, the dust particle distributions in the $(r, \phi)$ plane for 200 $\mu$m, 3.2 mm, and 6.4 mm dust sizes in Model 3, respectively. The black lines show radial density distribution of the gas normalized between (0, 2$\pi$), while the filled green circles mark the position of the planets. The width of the density profile is multiplied by a constant for readability.}
    \label{model3_dust}
\end{figure*}
\begin{figure}[h]
    \centering
    \includegraphics[width=\linewidth]{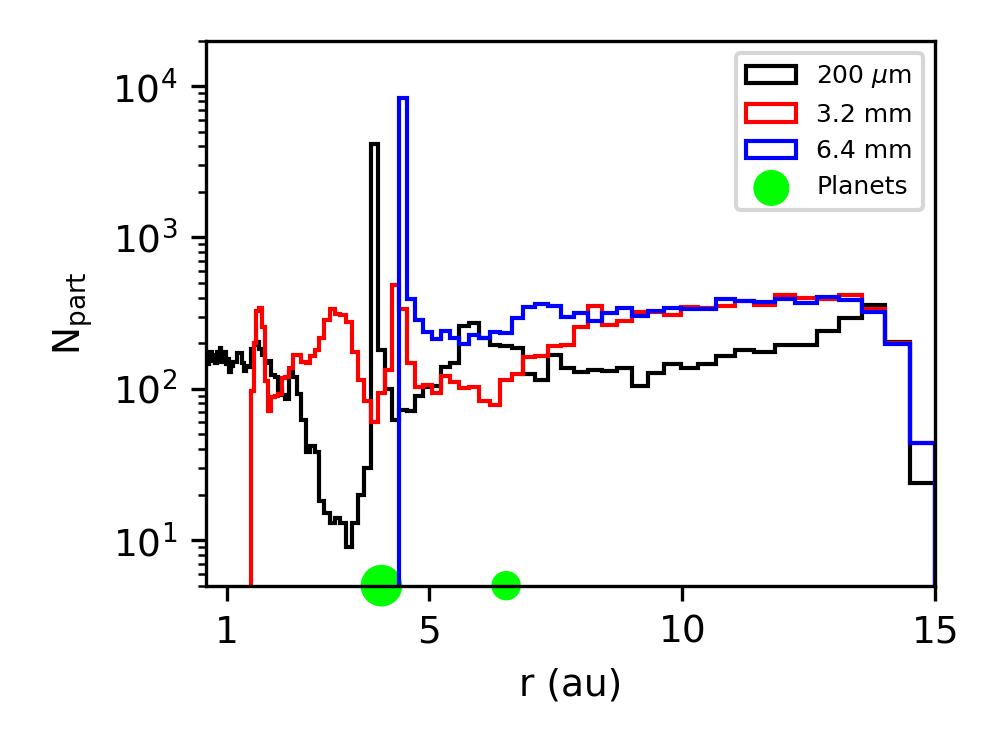}
    \caption{Same as Fig. \ref{model1_hist}, but for Model 3. The black line represents 200-micron particles, while the red and blue lines represent 3.2-mm and 6.4-mm grains, respectively.}
    \label{model3_hist}
\end{figure}
\begin{figure*}[h]
    \centering
    \includegraphics[width=\linewidth]{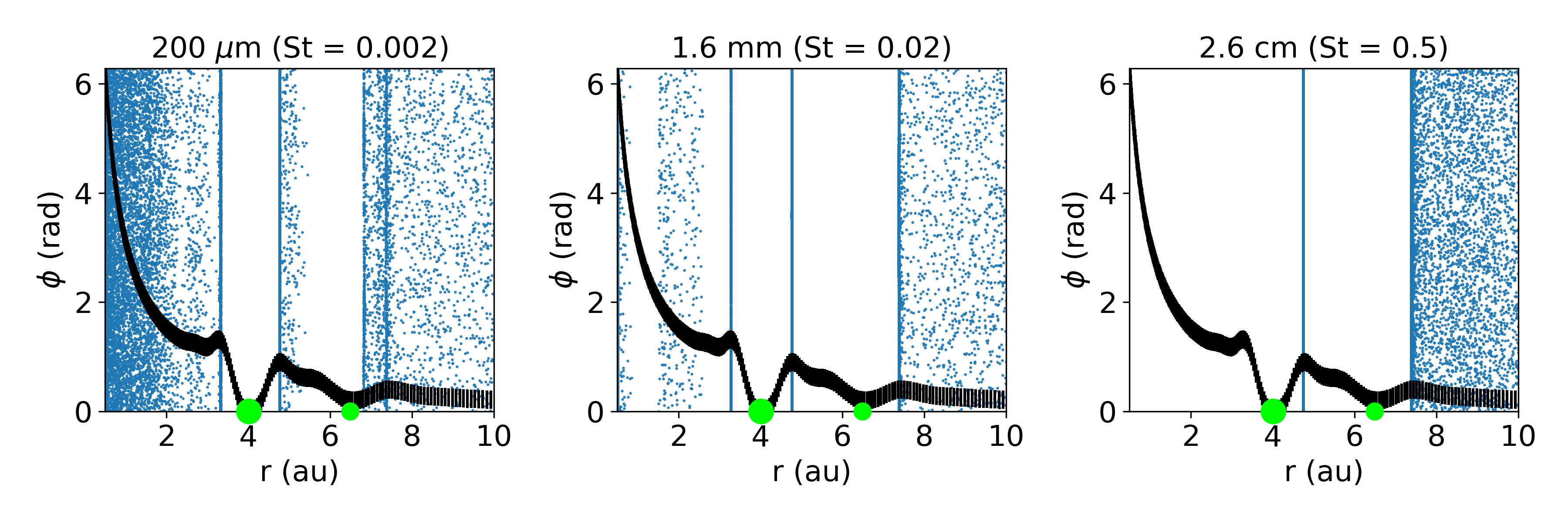}
    \caption{Dust distribution in Model 4. From left to right, the dust particle distributions in the $(r, \phi)$ plane for 200 $\mu$m, 1.6 mm, and 2.6 cm dust sizes in Model 4, respectively. The black lines show radial density distribution of the gas normalized between (0, 2$\pi$), while the filled green circles mark the position of the planets. The plots are zoomed in the inner region (0.5,10) au and the width of the density profile is multiplied by a constant for readability.}
    \label{model4_dust}
\end{figure*}
\begin{figure}[h]
    \centering
    \includegraphics[width=\linewidth]{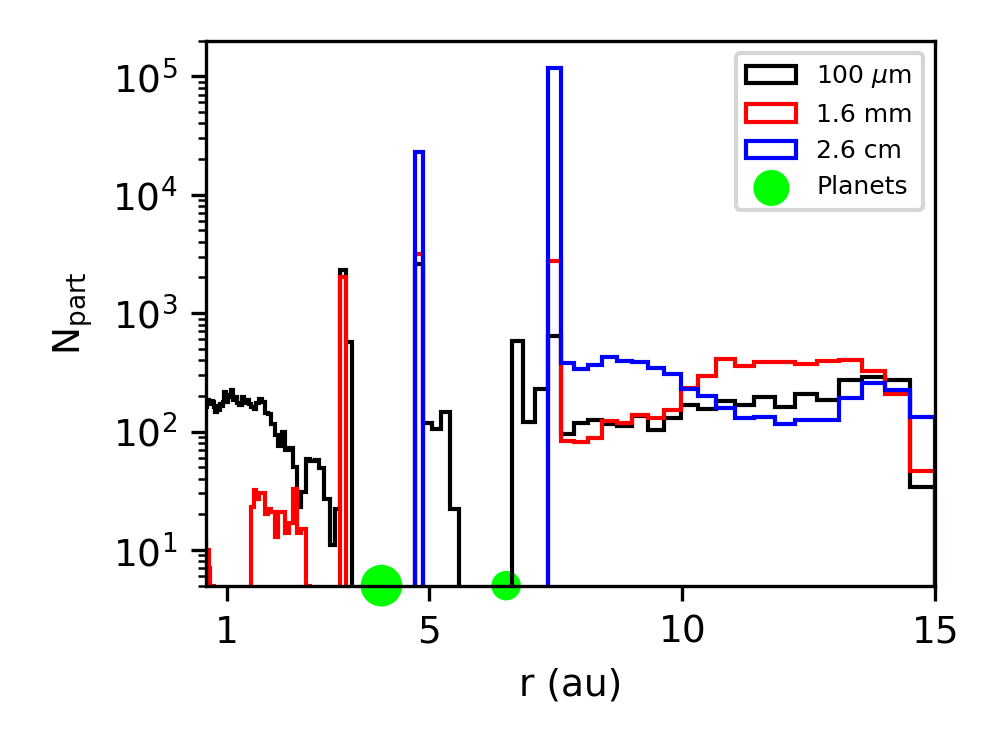}
    \caption{{Same as Fig. \ref{model1_hist}, but for Model 4}. The black line represents 100-micron particles, while the red and blue lines represent 1.6-mm and 2.6-cm grains, respectively.}
    \label{model4_hist}
\end{figure}
\begin{figure*}[h]
    \centering
    \includegraphics[width=\linewidth]{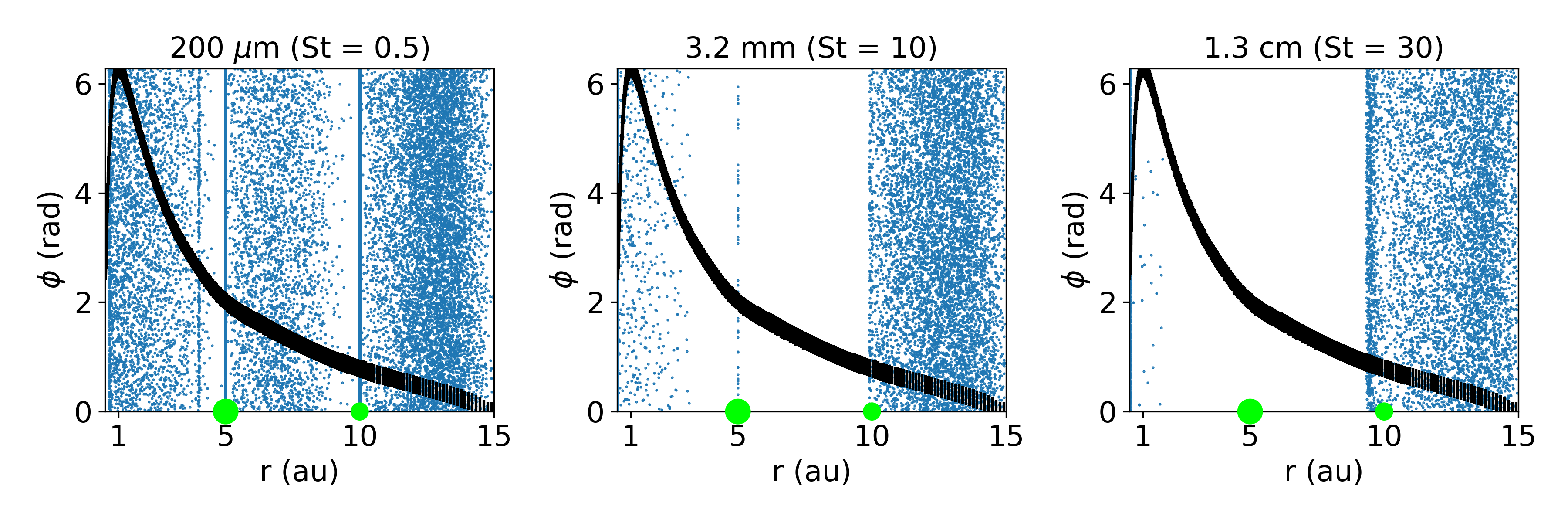}
    \caption{Dust distribution in Model 5. From left to right, the dust particle distributions in the $(r, \phi)$ plane for 200 $\mu$m, 3.2 mm, and 1.3 cm dust sizes in Model 5, respectively. The black lines show radial density distribution of the gas normalized between (0, 2$\pi$), while the filled green circles mark the position of the planets. The width of the density profile is multiplied by a constant for readability.}
    \label{model5_dust}
\end{figure*}
\begin{figure}[h]
    \centering
    \includegraphics[width=\linewidth]{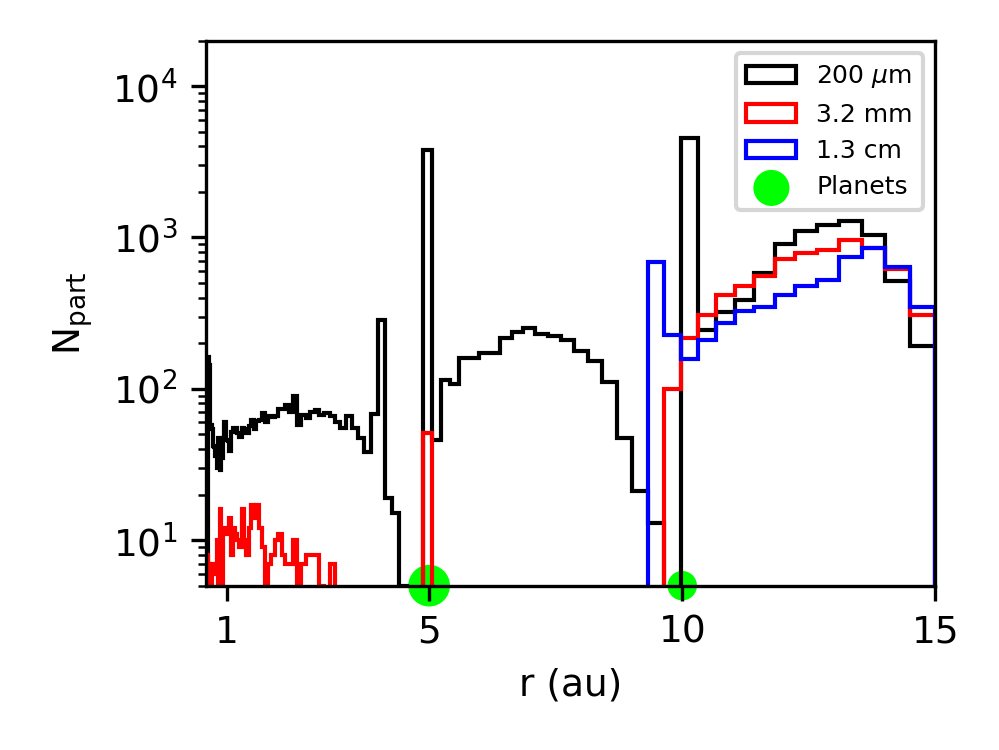}
    \caption{{Same as Fig. \ref{model1_hist}, but for Model 5. The black line represents 200-micron particles, while the red and blue lines represent 3.2-mm and 1.3-cm grains, respectively.}}
    \label{model5_hist}
\end{figure}
According to the mechanism proposed by \citet{2017MNRAS.469.1932D}, a low-mass planet can open a dust gap without perturbing the gas if the tidal torque it exerts on the dust particles overcomes the competing aerodynamic drag of the gas. This effect is most pronounced for weakly coupled particles, i.e., those with high Stokes numbers, where gas drag is relatively inefficient. To investigate this regime, we constructed Model 3, which adopts disk parameters that promote higher Stokes numbers. \\
Specifically, we considered a thinner and colder disk with a lower gas surface density of $\Sigma_0 = 100$ g/cm$^2$ (with a radial slope $p = 1.0$) and a temperature of $T_0 = 200$ K (with $q = 1.0$). The temperature gradient equal to 1 produces a geometrically flat disk; this has important implications for gap formation: in particular, a planet is expected to affect the local gas profile and open a dust gap if its mass exceeds the threshold mass given by $M_{\rm gap} \gtrsim 0.1 \, M_{\rm th}$ (\cite{2016MNRAS.459.2790R},\cite{2017MNRAS.469.1932D}). In flat disks, this threshold does not vary with the orbital radius, simplifying the analysis of the gap opening criteria across the disk. \\
In this model, we introduced an inner planet with $m_1 = 10 \, M_\oplus$ at $a_1 = 4$ au and an outer planet with $m_2 = 5 \, M_\oplus$ at $a_2 = 6.5$ au. Due to the low gas density, the migration torques are weak, and the planets remain nearly fixed in their orbits throughout the simulation (see Fig. \ref{orbital_evolution}). \\
The resulting gas and dust distributions for three representative grain sizes are shown in Figure \ref{model3_dust}. The first result is that the outer planet is virtually invisible in the gas and dust distributions. Its mass is insufficient to perturb the gas and it produces only shallow depletions of millimeter- and centimeter-size dust around its orbit. \\
In contrast, the inner planet has a more noticeable effect. Although not yet massive enough to carve a full gas gap or produce a pressure bump, it does induce a localized decrease in gas surface density near its orbit. This weak gas perturbation, combined with the tidal torque on high-Stokes-number particles, is sufficient to stop the inward drift of large grains ($s \gtrsim 6.4$ mm) at the outer edge of the planet’s orbit, consistent with the dust-only gap opening mechanism proposed by \citet{2017MNRAS.469.1932D}. \\
An important distinction from previous models is that a full dust cavity is formed inside the inner planet’s orbit. This arises because (i) large grains from the outer disk are halted at the orbit of the inner planet, and (ii) grains originally interior to the planet have already drifted inward and accreted onto the star. As a result, the region inside 4 au becomes depleted of large grains, producing a dust cavity rather than a localized gap, or a so-called transition disk. \\
To provide a more quantitative description of the substructures, we show in Fig. \ref{model3_hist} the number density of particles in Model 3 as a function of radial distance (see also Fig. \ref{gapwidths}). For the 200$\mu$m grains, we observe a dust gap around the inner planet with $\Delta _{\rm{gap}} \simeq 2$ au and a depth of about an order of magnitude in dust density. Note that the dust ring at the location of the inner planet is an artifact, as the dust particles coming close to the planet become trapped and would be accreted. Still, the dust depletion around the planet may imply that the gas outflow mechanism is playing a role even in this case because of the low Stokes number of these particles. For the 3.2-mm grains, we observe shallow depletions around both planets and a cavity inside 1 au, which has no clear interpretation: it could be related to the water snowline that is present at 1.1 au in this model, but it would be the only case in which the snowline produced significant dust features. For the 6.4-mm grains, a dust cavity is formed with $\Delta _{\rm{gap}} \simeq 4$ au, together with a dust ring at the outer edge of the inner planet's orbit, where the dust density is about 2 orders of magnitude above the average. \\
Due to the continuous influx of dust grains from the outer disk boundary in our simulations, large particles accumulated at the outer edge of the dust-depleted region, forming a pronounced dust ring. This pile-up could enhance local solid densities to the point where streaming instability is triggered, potentially facilitating further planetesimal or planet formation. Meanwhile, large grains originally located inside the dust cavity have been completely lost to inward drift, further highlighting the planet's filtering role in shaping the dust architecture.\\
To explore the impact of disk viscosity on dust gap formation, we performed an additional simulation, called Model 4, in which we adopted the same disk and planetary parameters as in Model 3, but significantly reduced the viscosity by setting $\alpha = 10^{-4}$. This model allowed us to isolate the effects of low-viscosity conditions on both the dynamical evolution of the planets and the resulting dust substructures. \\
In standard viscous accretion theory, the angular momentum is transported outward through turbulent stresses parameterized by $\alpha$ viscosity \citep{1974MNRAS.168..603L}. However, recent work suggests that angular momentum transport in some disks may be dominated by magnetized winds \citep[and references therein]{2023NewAR..9601674R}. In such low-viscosity environments, the dynamical coupling between planets and the gas disk is fundamentally altered. In fact, to open a gas gap, the planetary torque must overcome the viscous diffusion that tends to close the gap \citep{1986ApJ...309..846L, 1993prpl.conf..749L, 2015A&A...584A.110P}. \\
In Model 4, an immediate consequence of low viscosity is that the disk feedback to the planets is significantly weakened, as the planets maintained basically fixed orbits (see Fig. \ref{orbital_evolution}). Crucially, the gravitational torque exerted by the planets on the gas becomes significantly more effective in this regime. With less viscous diffusion to smooth out perturbations, even sub-thermal mass planets can carve pressure bumps in the gas distribution. These pressure maxima act as dust traps, halting the radial drift of grains of all sizes. Figures \ref{model4_dust} and \ref{model4_hist} clearly show the emergence of deep dust gaps around the planetary orbits, with the largest grains forming an extended cavity, similarly to Model 3. In particular, the left panel of Figure \ref{model4_dust} shows 200 $\mu$m grains forming two gaps around the orbits of the planets, separated by a ring. Intermediate-size grains (the central panel of Fig. \ref{model4_dust}) behave in a similar way, except for the inner region, which is more depleted of dust grains because they have already drifted toward the star. This is more evident for large grains (on the right panel), with almost all of them stopping at the outer edge of the outer planet's orbit, creating a transition disk. A key difference from Model 3 is that, in this low-viscosity scenario, dust accumulated at both edges of the pressure bump, forming sharp double-ringed structures in the dust distribution. This accumulation of particles can be clearly seen from the number density of dust in Figure \ref{model4_hist}. The dust gaps are narrower for small grains ($\Delta _{\mathrm{gap}} < 1.5$ au for 200-$\mu$m) and become progressively larger as the dust size increases ($\Delta _{\rm{gap}} \simeq 4$ au for the 2.6-cm size dust cavity, see also Fig. \ref{gapwidths}), while the dust rings are extremely narrow but reach a density that is two orders of magnitude higher than the initial one. This morphological signature, a central gap flanked by two narrow rings, is a direct consequence of the structure of the pressure bump in the gas and is not seen in the other gap opening mechanisms of Models 1–3. This provides a potentially powerful observational diagnostic: the detection of double rings in (sub)millimeter continuum images may indicate the presence of a pressure bump formed by a low-mass planet in a low-viscosity disk. \\
Moreover, we see in the central panel of Fig. \ref{model4_dust} and in Fig. \ref{model4_hist} that the distribution of 1.6-mm size grains presents secondary and tertiary gaps interior to the orbit of the inner planet, around 3 and 1 au, respectively. The formation of multiple gaps by a super-Earth in a low-viscosity disk has been studied by \cite{2017ApJ...850..201B} and they related it to the spiral arms produced by the planet in the gas distribution, which steepen into shocks (see also \cite{2018ApJ...866..110D}). The fact that we observe multiple gaps produced by a planet only in a low-viscosity scenario is in agreement with previous results \citep{2017ApJ...850..201B}.\\
The comparison between Models 3 and 4 underscores the crucial role of disk viscosity in shaping both gas and dust structures for a given planetary system. In viscous disks, feedback onto the planets is stronger, which makes it hard for them to perturb the gas distribution, but they can still open gaps in the dust. In contrast, low-viscosity disks allow for the formation of long-lived pressure bumps that, in turn, trap dust efficiently and produce observable substructures that are correlated with gas features. These results show that the interpretation of dust gaps in circumstellar disks requires careful consideration of the viscosity of the disk. The same planetary system can generate drastically different observable features depending on the value of $\alpha$. Therefore, combining morphological analysis of dust rings with constraints on disk turbulence, for example from gas kinematics, may be essential to accurately infer the presence and properties of embedded planets. \\
In Model 5, we considered two planets with masses of $m_1 = 10$ M$_{\oplus}$ and $m_2 = 5$ M$_{\oplus}$ and semimajor axes of $a_1 = 5$ au and $a_2 = 10$ au. The disk parameters are the same as in the simulations performed by \cite{2017MNRAS.469.1932D}: the gas density is $\Sigma _0$ = 0.1 g/cm$^2$ at 1 au with the power law index $p = 0.1$ and the disk temperature is $T_0 = 300$ K with the index $q = 0.5$. In such a low-density disk, the dust particles are weakly coupled to the gas, so that the Stokes number is higher and the gravitational interaction with the planets is stronger. Moreover, a low gas density implies very slow migration, so it has no impact on the gap opening process (see Fig. \ref{orbital_evolution}). In this configuration, the mass of the planets is below $0.1 M_{th}$, so they cannot create a gas pressure bump in the disk. However, they are still perturbing the dust distribution, as shown in Figure \ref{model5_dust}. In particular, most of the 200-$\mu$m grains (on the left in the figure) have been able to pass through the orbits of the planets, while a gap has opened in between the two orbits for the 3.2 mm grains (at the center of Fig. \ref{model5_dust}) and all the 1.3 centimeter-size grains (on the right) have stopped at the location of the outer planet, leaving an inner cavity. We quantify the substructures showing the density of dust particles as a function of the radial distance in Figure \ref{model5_hist}. The 200 $\mu$m size grains form two gaps ($\Delta _{\mathrm{gap}} \simeq 1$ au, see Fig. \ref{gapwidths}) around the orbit of the planets, accompanied by two rings with density about 1.5 orders of magnitude above the average. The 3.2-millimeter-size grains show a wider interior gap ($\Delta _{\mathrm{gap}} \simeq 8$ au) to the orbit of the outer planet, with smaller concentrations of dust at the location of the planets, which are dust particles that would have been accreted. Finally, the 1.3-cm grains present a cavity with a width of $\Delta \simeq 9$ au interior to the orbit of the outer planet. In summary, a single gap (or cavity) is created by two planets when the outer one is stopping the grains from drifting inside, while the dust in the inner region has already drifted toward the star. This shows that the dust distribution in a multi-planet system is not the outcome expected by the sum of the contributions of isolated planets, as multiple bodies may be hidden inside a single dust gap.
\subsection{Single planet cases}
\begin{figure}[h!]
    \centering
    \includegraphics[width=\linewidth]{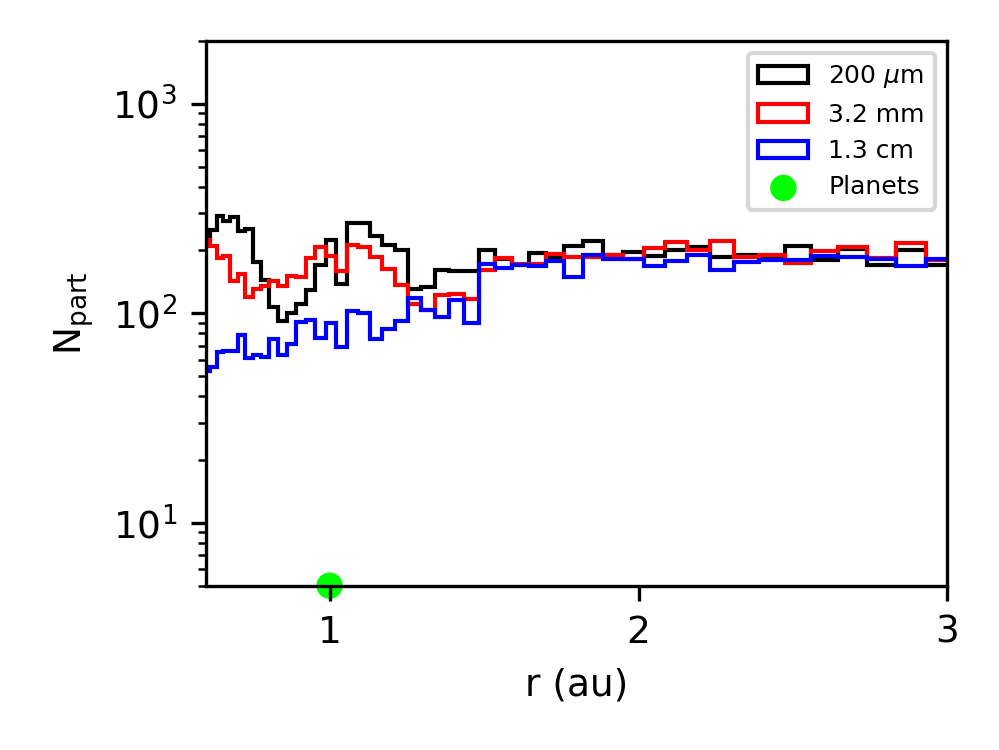}
    \caption{Same as Fig. \ref{model2_hist}, but for the case of Model 2S.}
    \label{single_mod2}
\end{figure}
\begin{figure}[h!]
    \centering
    \includegraphics[width=\linewidth]{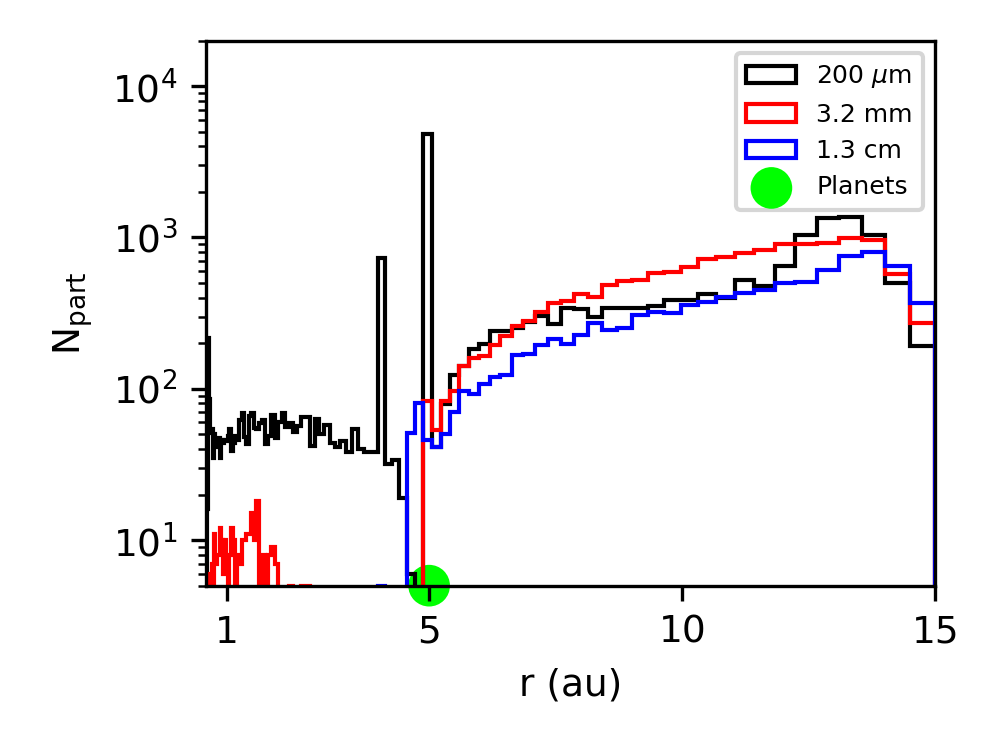}
    \caption{Same as Fig. \ref{model5_hist}, but for the case of Model 5S.}
    \label{single_mod5}
\end{figure}
We have claimed that multi-planet systems perturb the dust distribution in a way that cannot be explained by considering each planet in isolation. To verify the robustness of this conclusion, we performed simulations with isolated planets under the same conditions as the previous models. In particular, we show in Figure \ref{single_mod2} the result of Model 2S, a simulation with the same disk parameters as Model 2 but with a single Earth-mass planet located at 1 au (that is, the first planet considered in Model 2), as an example of models with low Stokes numbers. In this case, the dust distribution presents only minor perturbations around the planet's orbit for the small grains, but no gap is observed at 2 au. This means that one Earth-mass planet is not enough to significantly perturb the dust component under these conditions, but the combined influence of two equal-mass planets can open a deep and narrow dust gap for small grains (as shown in Figure \ref{model2_hist}). \\
Similarly, in Figure \ref{single_mod5} we show the result of Model 5S, a simulation with the same disk parameters of Model 5 but with a single planet with mass $m = 10$ M$_{\oplus}$ located at 5 au, which is the first planet in Model 5, as an example of models with high Stokes numbers. In this case, the planet can open a gap for millimeter- and centimeter-size grains, but this gap is only interior to the orbit of the planet. This is a key difference from Fig. \ref{model5_hist}, where the presence of a lower-mass (5 M$_{\oplus}$) planet at 10 au is enough to open an extended gap in between the orbits of the two planets. This common gap may be misinterpreted in observations as due to a single planet located in the middle of the gap, but our results show that this feature can only be explained by invoking the presence of multiple planets.
\subsection{Radial gap widths}
\begin{figure*}[h!]
    \centering
    \includegraphics[width=\linewidth]{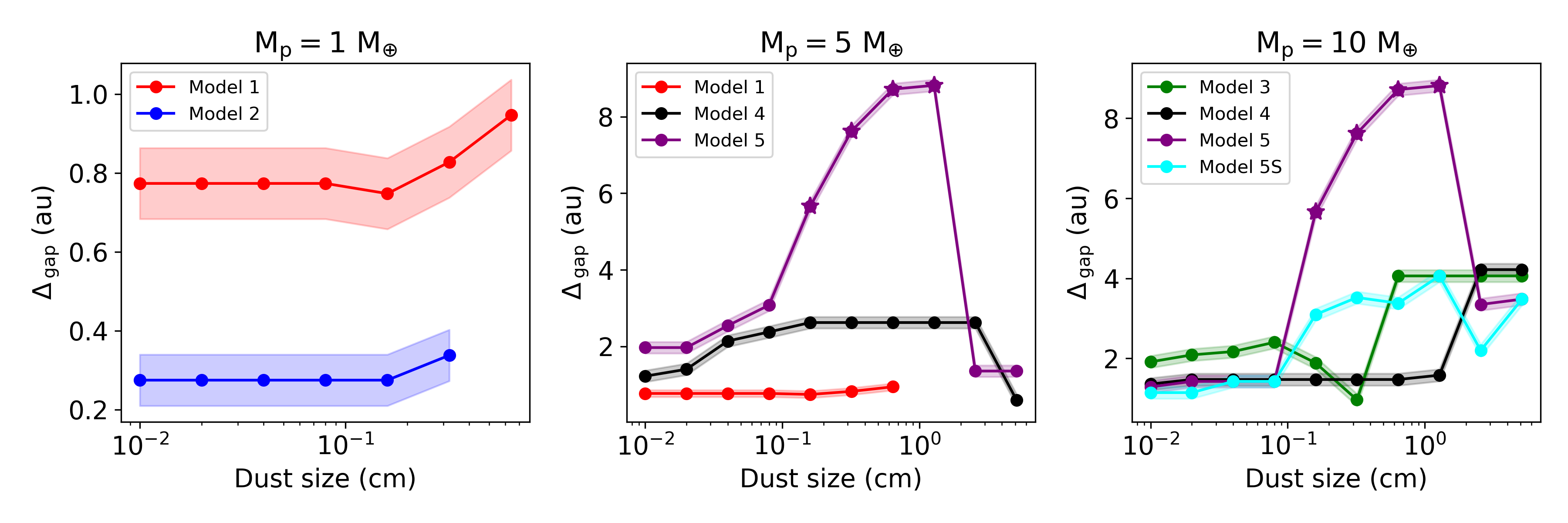}
    \caption{Radial gap widths. From left to right, radial gap width as a function of dust size for three different planetary masses (1, 5, 10 $M_{\oplus}$). The shaded regions represent the uncertainty in the width measurement, which is equal to the local width of the bins used for computing the dust density. Points denoted with stars refer to gaps common to both planets, so they are reported for each one. When no gap is observed (e.g., for the outer planet in Model 3 and the single planet in Model 2S), no width is reported.}
    \label{gapwidths}
\end{figure*}
To provide a more quantitative analysis, we computed the radial gap width for each model. First, we divided the dust distribution into 100 radial bins with logarithmic spacing (similar to the simulation grid) to calculate the 1D dust density profile $\Sigma _{\rm{D}} (r)$ as a function of the radial distance from the star. Then we normalized this profile by the initial density, $\Sigma _{\rm{D},0}$: the inner and outer radius of the gap are found where this normalized density exceeds a threshold of $\Sigma _{\rm{D,norm}} > f = 0.1$. If a dust peak were found at the location of the planet, we neglected it because it is related to dust particles that are trapped in the horseshoe region and would be accreted. In the case of a dust cavity, the width is given by the distance between the first density peak and the internal boundary of the simulation grid, which is 0.5 au. This method of measuring the width of the gap differs from that employed in previous work (for example, \cite{2017ApJ...835..146D}), which considered the distance between the points where the dust density reaches the geometric mean of the initial and final densities in the center of the gap. This does not apply in our case because for many dust sizes the density is 0 inside the gap (for the same reason, we cannot give an accurate measurement of the gap depth). \\
In Figure \ref{gapwidths} we show the measured gap width as a function of dust size for each model, focusing only on the primary gap for each planet. We associated an uncertainty with each measurement given by the local width of the bins. In the high-St models, the gaps are significantly wider, indicating that the tidal torque mechanism may be more efficient in opening dust gaps, but requires higher planet masses compared to the low-St models. On average, the width of the gap increases with dust size, with the largest widths measured for the common gaps observed in Model 5. However, a common gap does not form for the 2.56 and 5.12 cm grains: this is probably because of the high drift speed of these grains, which migrate past the outer planet, causing a fast influx of dust grains and forming a transition-type disk, rather than a common gap. In the low-St models (1 and 2), the dust gaps are much narrower, with $\Delta _{\rm{gap}} < 1$ au, and form only for the smallest grains. This is because the gas outflow mechanism is expected to be efficient only for low-St particles in the vicinity of the planet's orbit. In addition, the outer planet of Model 3 and the single planet in Model 2S show shallow dust depletions (see Figs. \ref{model3_hist}, \ref{single_mod2}) that are below the gap threshold (i.e., less than an order of magnitude compared to the initial density), so they are not reported in Figure \ref{gapwidths}. In summary, the presence of multiple planets can have different effects: they may open a dust gap where a single planet may not be able to (Models 2, 2S); they may shift the location of the gap inside or outside with respect to the planets' orbit (Models 1 and 2); they may open a common gap which is significantly more extended compared to single gaps generated by isolated planets (Models 3, 5, 5S); in some cases, they may create multiple gaps and rings (Model 4).\\ 
\section{Discussion}
Our results on the formation of dust substructures by sub-thermal mass planets present some similarities with previous work. \cite{2013ApJ...769...41D} studied the opening of a gas gap by a Neptune-mass planet ($q \sim 10^{-4}$) in the gas distribution of a low-viscosity disk ($\alpha < 10^{-4}$), and found no minimum mass limit for a planet to open a gas gap in the low-viscosity limit. This result is in agreement with what we find in Model 4, where we studied the gap opening process by super-Earths in a low-viscosity regime. By incorporating the dust component into the analysis, we were able to characterize the potentially observable dust substructures produced by the planets in this regime, even when the planet's Hill radius is lower than the disk scale height (i.e., the planet has a sub-thermal mass).
\subsection{Multiple gas gaps and their origin} 
We observed in Model 4 that low-mass planets can actually open multiple dust gaps, in particular for millimeter-size grains (see Figures \ref{model4_dust} and \ref{model4_hist}). These multiple gaps could be related to spiral arms generated by the planet in the gas distribution, which steepen into shocks \citep{2017ApJ...850..201B, 2018ApJ...866..110D, 2018ApJ...859..118B, 2021ApJ...912...56B}. However, we do not find significant perturbations in the gas beyond the first gap in our analysis (as seen from the radial density profile in Figure \ref{model4_dust}). In contrast, secondary and tertiary dust gaps are observed only interior to the orbit of the inner planet in the millimeter-sized grains, which may imply that the gap opening process is size-dependent and is much more effective in the dust than in the gas. Moreover, according to \cite{2017ApJ...850..201B} multiple gas gaps driven by spiral shocks are generated only if the disk viscosity is sufficiently low ($\alpha < 5 \times 10^{-4}$), so this mechanism cannot reproduce the result we found in Model 2, where $\alpha = 4 \times 10^{-3}$ and the dust substructures are found to overlap with the Lindblad resonances of the two planets. Another potential relationship between dust gaps and Lindblad resonances could be in Model 1, where the gap is close to the first outer resonance for the inner, most massive planet. We note that each of the models presented explores a different region of the parameter space: as a consequence, different physical mechanisms can dominate, which are not easily separable. Still, the difference we found between the dust distribution produced by a single planet compared to a multi-planet system can also be seen in the work by \cite{2015ApJ...809...93D}: they observe that the gap produced by a single, Jupiter-mass planet has a width that is 45\% of the one produced by four planets of equal mass. This implies that the dust distribution generated by a multiple system cannot be reproduced simply by summing up the contributions of single planets. \\
Moreover, our results may have some implications for observational findings in multi-gapped and transition disks. Importantly, recent statistical analysis of the molecules with ALMA at planet-forming scales (MAPS) survey has shown no significant correlation between molecular line emission and continuum substructures, challenging the usual pressure bump mechanism for gap opening \citep{2022ApJ...924L..31J}. Note that pressure substructures may still be present, since optical depths and chemical abundances also play major roles in shaping the observed line emission. \cite{2023A&A...674A.113I} found correlations between negative rotational velocity gradients (tracing pressure bumps) and dust substructures in most MAPS sources.
Nevertheless, if dust substructures are observed without correlated gas perturbations, they can be interpreted as due to radial outflows or tidal torques produced by low-mass planets, depending on disk parameters, as we have shown in our analysis. Conversely, when a dust substructure is found to be correlated with gas pressure bumps, our results suggest that either the planet is super-thermal (M $\gtrsim$ M$_{\rm{th}}$), or the disk viscosity is low ($\alpha \lesssim 10 ^{-4}$). \\
In addition to rings and gaps, nonaxisymmetric dust features could be produced by planet-disk interactions, such as crescent-shaped asymmetries due to dust trapping at Lagrange points \citep{2021A&A...647A.174R}. We do not observe such features in any of our models, probably because of the modest mass of the planets considered.
\subsection{TW Hya case}
Multiple bright and dark dust rings have been observed from 1 to 47 au \citep{2021A&A...648A..33M}. These rings are too narrow and shallow to be produced by giant planets, but we have shown that smaller bodies are possible. In particular, bright and dark rings in the distribution of micrometer- and millimeter-sized grains could be produced by multiple low-mass planets, like the ones we show in Models 1 and 2, which are not correlated with gas perturbations (see Figures \ref{model1_dust} and \ref{model2_dust}).
\subsection{J1604 case} 
In this transition disk with a prominent dust ring observed at 81 au, the surface density of the gas exhibits a drop at 60 au and a pressure maximum that matches the location of the dust ring \citep{2025ApJ...984L..19Y}, while the viscosity of the disk is found to be as low as $\alpha \sim 2 \times 10^{-4}$. With this low viscosity value, the mass of the putative planet responsible for the dust cavity may be significantly lower than previously expected, based on our results in Model 4. In fact, at a distance of 60 au, the Stokes number of dust particles would be much higher than in our configuration, and the resulting dust distribution could resemble the right panel of Figure \ref{model4_dust}, where the planets generate an inner dust cavity limited by a prominent dust ring.
\subsection{HL Tau case} 
One of the dust gaps is found to be correlated with a drop in the column density of HCO$^+$ \citep{2019ApJ...880...69Y}, suggesting that the pressure bump mechanism is the dominant one. This implies that the mass of the putative planets would be around 0.2-0.5 $M_{J}$, as was shown in previous works \citep{2015MNRAS.453L..73D, 2015ApJ...809...93D}, much higher than the masses of the planets that we considered.
\subsection{Caveats} \label{caveats}
As was mentioned before, we did not include dust evolution in our simulations. By dust evolution, we mean the introduction of a realistic size distribution which can, for example, be tuned to follow a power law. The variation in the number of particles within each bin during the evolution of the system can then be translated into a modification of the power-law size distribution. However, this approach does not account for dust particle growth, which may be more relevant for determining the dust size distribution. For this reason, we limited our study to the dynamics of dust particles. \\
Moreover, we adopted a 2D, locally isothermal setup. The omission of heating and cooling processes and the full vertical structure may prevent the development of additional instabilities, such as multiple spiral arms \citep{2021ApJ...912...56B}. However, a 3D setup would significantly increase computational cost, especially for simulations including a large number of Lagrangian particles with a wide size distribution, and could introduce numerical instabilities. By removing vertical and thermal complexity, we can clearly track substructure formation for individual dust sizes and maintain a stable, reproducible framework for multi-size dust populations, serving as a baseline for understanding gap opening mechanisms before including full 3D structure or radiative effects in future work.
\section{Conclusions}
In this paper, we have investigated the formation of dust gaps in circumstellar disks driven by the presence of multiple sub-thermal mass planets, using 2D hydrodynamical simulations with dust treated as Lagrangian particles. Here we summarize the main results:
\begin{itemize}
    \item Dust gaps can open due to radial gas outflows (for tightly coupled particles) and tidal torques (for weakly coupled particles), without requiring perturbations in the gas pressure gradient. Our parameter space analysis shows that the gas outflow mechanism dominates for low Stokes numbers (St < $10^{-2}$), whereas tidal torques dominate for St > $1$; in the intermediate regime ($10^{-2} \lesssim $ St $\lesssim 1 $), no gaps are observed for planet masses M$_p$ < 10 M$_{\oplus}$, suggesting that no gap opening mechanism is at work in this case;
    \item Our models with low Stokes numbers show that sub-thermal mass planets can open narrow dust gaps ($\Delta _{\rm{gap}} \lesssim 1$ au; see Fig. \ref{gapwidths}) without any correlated gas structures. This is consistent with the outflow mechanism proposed by \cite{2022A&A...665A.122K} for a single planet. However, in multiple systems, we find that dust gaps may be displaced with respect to planetary orbits, particularly in dynamically evolving systems with migrating planets, where we also find gaps that overlap with Lindblad resonances (see Fig. \ref{model2_sig});
    \item Our models for high Stokes numbers show that sub-thermal mass planets open large dust gaps and cavities ($\Delta _{\rm{gap}} \gtrsim 4$ au) without gas pressure bumps, consistent with the tidal torque mechanism proposed by \cite{2017MNRAS.469.1932D}. However, the dust morphology can change significantly when two or more planets are present in the system. Multiple planets can open a gap where a single planet is not able to, or a common gap may form (see Fig. \ref{gapwidths}); as a consequence, estimating the planet mass in observed disks assuming that it acts alone may lead to inaccurate conclusions. We plan to extend this work to super-thermal mass planets in a future paper (Marzari et al., in prep.);
    \item Our findings indicate that the physical properties of the disk are key
    in determining the dominant gap opening mechanism within a system. This
    degeneracy introduces a significant obstacle to interpreting observed
    dust substructures as direct signatures of planet formation.
\end{itemize} 

\begin{acknowledgements}
This publication was produced while attending the PhD program in Astronomy at the University of Padova, Cycle XXXIX, with the support of a scholarship co-financed by the Ministerial Decree no. 118 of 2nd March 2023,760 based on the NRRP - funded by the European Union - NextGenerationEU - Mission 4 Component 1 – CUP C96E23000340001. We thank the anonymous referee for their constructive comments, which helped to improve the quality of this paper.
\end{acknowledgements}

\bibliographystyle{aa}
\bibliography{references.bib}

\begin{appendix}
\section{Planet migration}
\begin{figure}[h!]
        \centering
        \includegraphics[width=\linewidth]{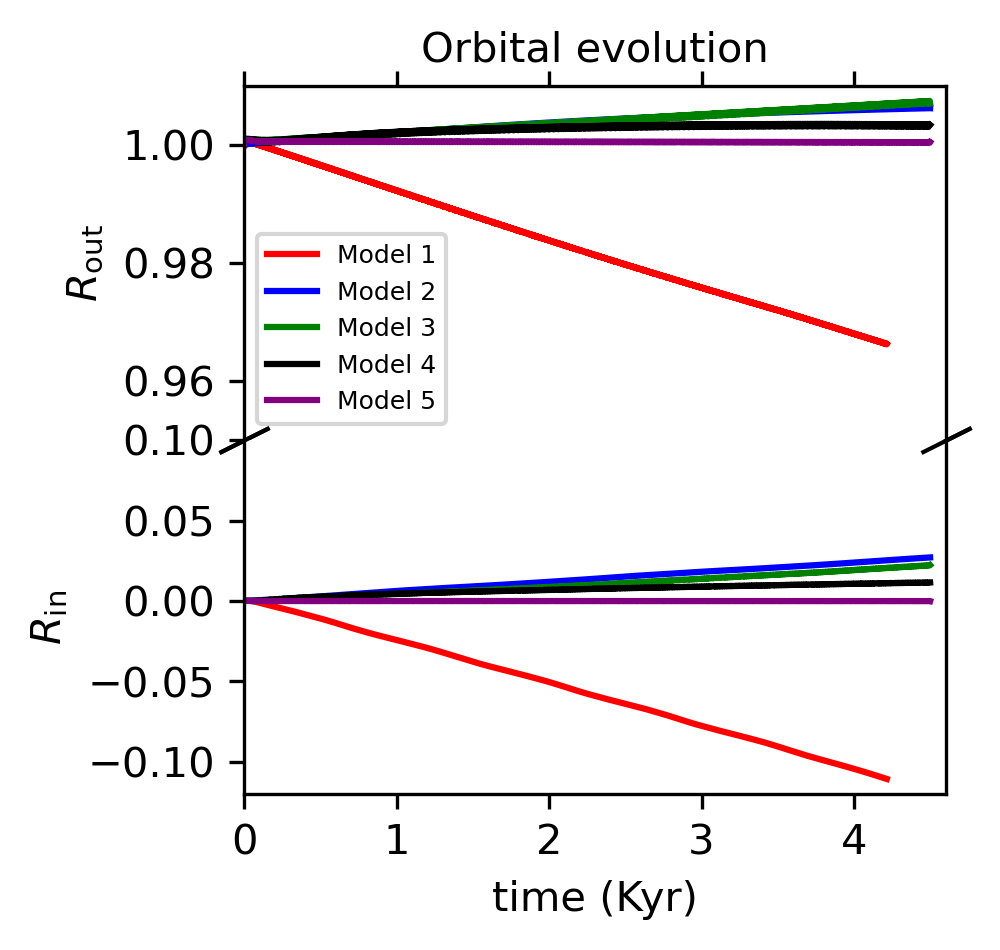}
        \caption{Orbital evolution of the planets across all models. The y axis is normalized so that $R_{\rm{out}}$ = 1 and $R_{\rm{in}}$ = 0 represent the outer and inner planets. The plot shows only the regions around 1 and 0 to highlight the differences between the models.}
        \label{orbital_evolution}
    \end{figure}

We show in Figure \ref{orbital_evolution} the orbital evolution of the planets in all models. Migration is generally slow because of the low mass of the planets, with the exception of Model 1 due to the high gas density and disk scale height. In all cases considered, planet migration still prevents the planets from being captured in the 2:1 MMR. The faster migration of Model 1 (and partially in Model 2) could affect the displacement of the dust gap with respect to the positions of the planets. In fact, in Model 1 both planets are migrating inward and the dust gap is located slightly inside the orbit of the outer planet, while in Model 2 both planets are slightly migrating outward and we observe the gap outside of the orbit of the outer planet. In general, planet migration is not expected to have a significant impact on whether the planets can form a gap or not in our simulations. In fact, given the low mass of the planets, the low gas surface density and the high gas temperature (compared to models where the planets are farther out), the migration timescale is significantly longer than the gap formation timescale in all our models.

\section{Dust particle eccentricity}
\begin{figure}[h!]
    \centering
    \includegraphics[width=\linewidth]{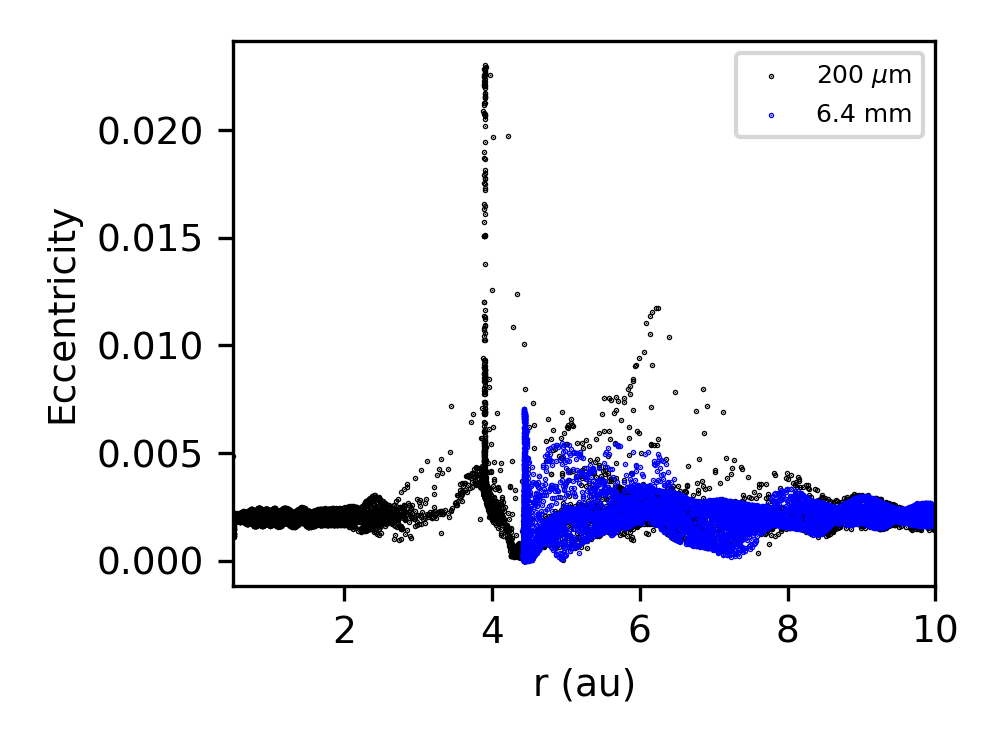}
    \caption{Orbital eccentricity of dust particles as a function of radial distance in Model 3. Black dots represent 200 $\mu$m grains, while blue dots correspond to 6.4 mm particles.}
    \label{model3_eccentricity}
\end{figure}

\begin{figure}[h!]
    \centering
    \includegraphics[width=\linewidth]{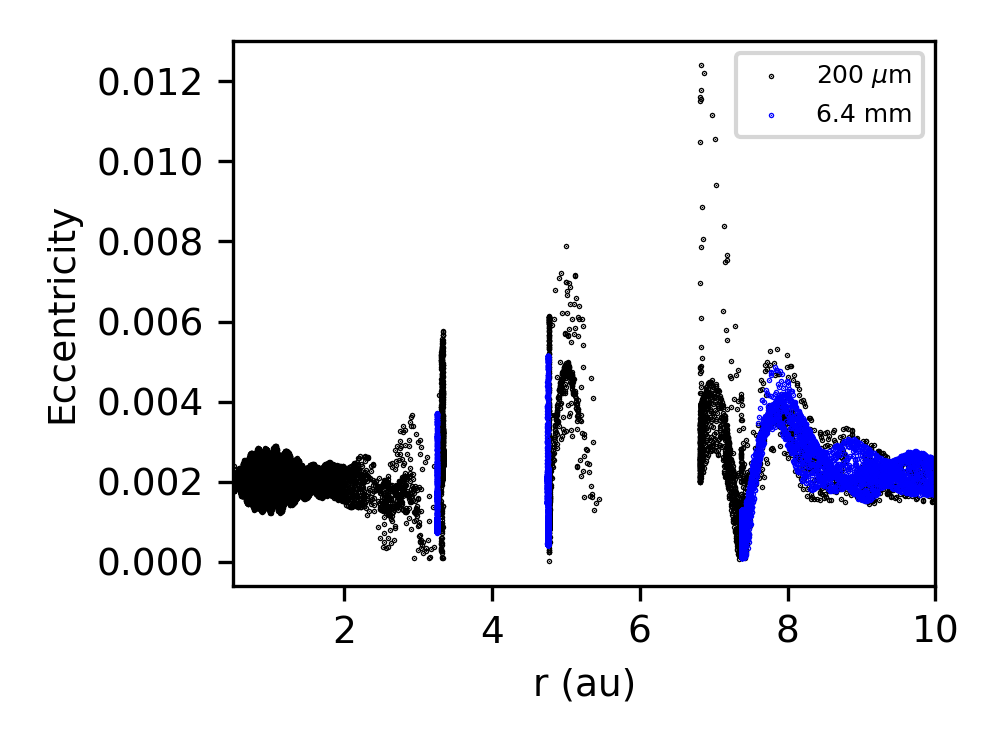}
    \caption{Same as Figure \ref{model3_eccentricity}, but for Model 4.}
    \label{model4_eccentricity}
\end{figure}

In addition to planet migration, the eccentricity of dust particles is another factor that can change the morphology of dust substructures. Overall, the eccentricities of dust grains remain extremely low across all our models, as expected given the modest planetary masses, which are not able to significantly excite the dust motion. The cases with the highest dust eccentricity are shown in Figures \ref{model3_eccentricity} and \ref{model4_eccentricity}, which highlight the orbital eccentricities of small and large dust grains in Model 3 and Model 4. We see that few particles with eccentricity e $\geq 0.01$ have accumulated near the most massive planet at 4 au - this behavior may be related to particles passing very close to the planet, which would have been otherwise accreted.

\end{appendix}

\end{document}